\documentclass[%
 reprint,
 superscriptaddress,
 amsmath,amssymb,
 aps,
 prb
]{revtex4-1}

\usepackage{braket}
\usepackage{slashed}
\usepackage{graphicx}% Include figure files
\usepackage{dcolumn}% Align table columns on decimal point
\usepackage{bm}% bold math
\usepackage{siunitx}
\usepackage{xcolor} % color text
\usepackage{tabularx}
\usepackage{dsfont}					%for the double-stroke font for the "identity matrix 1"
\usepackage{mathrsfs}				%for the script lettering
\usepackage{float}
\usepackage{hyperref}% add hypertext capabilities
\newcommand{\comment}[1]{}	%does not typeset text contained in argument
\newcommand{\tr}[0]{\mathbf{tr}}	%generates formatting for trace function 
\newcommand{\Tr}[0]{\mathbf{Tr}}	%generates formatting for trace function 
\newcommand{\Det}[0]{\mathbf{Det}} %generates formatting for determinant with capital

\newcommand{\be}[0]{\begin{equation}}	%Short hand for equation environment
\newcommand{\ee}[0]{\end{equation}}

\usepackage{hyperref}
\hypersetup{colorlinks, 
	linkcolor={blue!75!black!80!yellow},
	citecolor={blue!75!black!80!yellow}, 
	urlcolor={blue!75!black!80!yellow}
	}

\usepackage[normalem]{ulem}

\begin{document}

%\preprint{}

\title{Finite-Momentum Instability of Dynamical Axion Insulator}


\author{Jonathan B. Curtis}
\affiliation{College of Letters and Science, University of California, Los Angeles CA 90095, USA}
\affiliation{John A. Paulson School of Applied Sciences and Engineering, Harvard University, Cambridge Massachusetts 02138 USA}
\email[]{joncurtis@ucla.edu}
\author{Ioannis Petrides}
\affiliation{College of Letters and Science, University of California, Los Angeles CA 90095, USA}
\affiliation{John A. Paulson School of Applied Sciences and Engineering, Harvard University, Cambridge Massachusetts 02138 USA}
\author{Prineha Narang}
\affiliation{College of Letters and Science, University of California, Los Angeles CA 90095, USA}
\affiliation{John A. Paulson School of Applied Sciences and Engineering, Harvard University, Cambridge Massachusetts 02138 USA}

\date{\today}

\begin{abstract}
Due to the chiral anomaly, Weyl semimetals can exhibit a signature topological magnetoelectric response known as an axion term which is determined by the microscopic band structure.
In the presence of strong interactions Weyl fermions may form a chiral condensate, with the intrinsic dynamics and fluctuations of the associated condensate phase producing a dynamical contirbution to the axion response.
Here we show that an imbalance in the density of right- and left-handed electrons drives an instability of the chiral condensate towards finite momentum and leads to strong fluctuations in the axion response.
We suggest a long-wavelength theory of Lifschitz type governing the dynamics of the Goldstone mode and use this to characterize its associated spatial fluctuations, which manifest as an inhomogeneous anomalous Hall effect.
We show that these fluctuations produce signatures in inelastic light scattering experiments across a broad spectrum of frequencies, and can be used to determine the structure factor for the axionic collective mode.
\end{abstract}

\maketitle

{\it Introduction.}\textemdash
One of the most fascinating developments in condensed matter physics has been uncovering the fundamental role that topology plays in quantum systems~\cite{Jackiw.1976,Haldane.1983,Thouless.1982,Laughlin.1983,Kosterlitz.1973,Berezinskii.1972,Narang.2021}.
A key idea in this framework is that of the chiral anomaly, which originally was found in the context of high-energy physics~\cite{Adler.1969,Bell.1969}, but is now understood to have an important role in condensed-matter systems~\cite{Else.2021,Wang.2021a9i}.
Qualitatively, the chiral anomaly occurs when the classical action has symmetries which are not compatible with the quantum partition function~\cite{Fujikawa.1979,Fujikawa.2004}, leading to a breakdown of the conservation laws typically guaranteed by Noether's theorem; in this case the symmetry is called ``anomalous." 
This seemingly abstract idea has direct observable consequences for Weyl semimetals~\cite{Yan.2016,Armitage.2018,Meng.2012,Burkov.2011,Zyuzin.2012,Zyuzin.2017,Zyuzin.2012cdp,Zeng.2022,Raines.2017,Burkov.2014}.

Weyl semimetals feature a low-energy effective description in terms of gapless spin-1/2 electrons which come in pairs of opposite chirality.
Chiral symmetry then leads to the conservation of particle number for each chirality separately, strongly constraining their hydrodynamic responses and giving rise to interesting topological effects.
Hence, understanding the fate of these systems in the face of strong interactions is critical~\cite{Rylands.2021,Rylands.2022,Maciejko.2014,Balatsky.1990,Srivatsa.2018,Roy.2017}, especially due to their potential for technological applications~\cite{Zhao.2020}.

\begin{figure*}
    \centering
    \includegraphics[width=\linewidth]{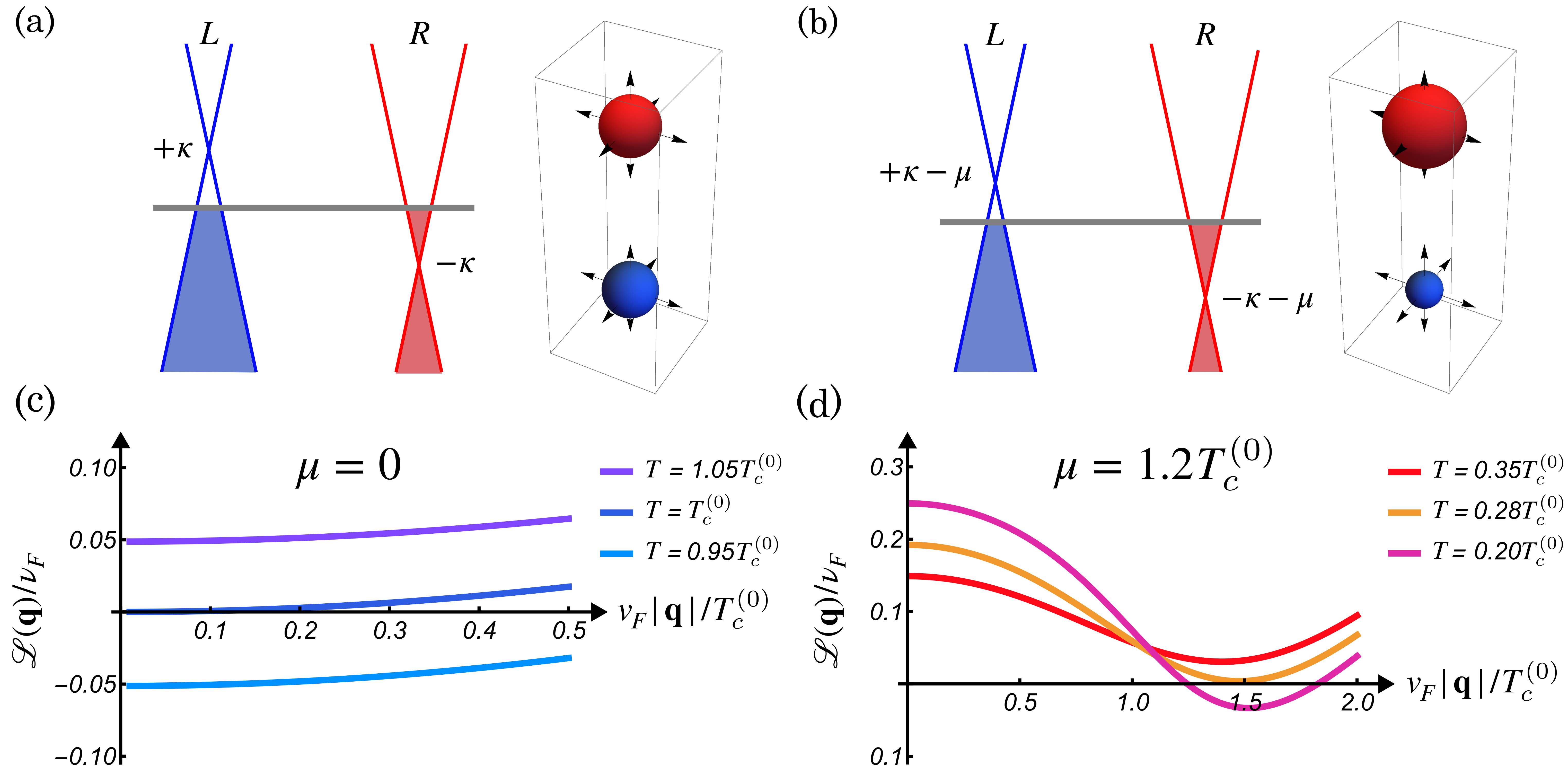}
    \caption{(a) Schematic depiction of Weyl fermions in the presence of $\kappa > 0$, with $\mu = 0$. 
    This has exactly compensated particle and hole pockets with right ($R$) and left ($L$) chiralities respectively, leading to equal sized Fermi surfaces of fixed helicity ${\bm \sigma }=\hat{\mathbf{p}}$.
    (b) Schematic depiction of Weyl fermions the presence of $\kappa > 0$, with $\mu = 1.2 T_c^{(0)}$. 
    This has uncompensated particle and hole pockets, leading to imbalanced Fermi surfaces.
    (c) Chiral condensate collective-mode susceptibility $\mathscr{L}(\mathbf{q})$ in the compensated $\mu = 0$ case for different temperatures as a function of $\mathbf{q}$. 
    The condensate forms at $T = T_c^{(0)}$ and at $\mathbf{q} = 0$.
    (d) Chiral condensate collective-mode susceptibility $\mathscr{L}(\mathbf{q})$ in the uncompensated $\mu = 1.2T_c^{(0)}$ case for different temperatures as a function of $\mathbf{q}$. 
    The condensate forms at much lower temperature $T_c < T_c^{(0)}$, and occurs at finite momentum.
    }
    \label{fig:main-figure}
\end{figure*}

Of particular interest is when interactions lead to a spontaneous breaking of the chiral symmetry.
This possibility was originally proposed in the celebrated Nambu-Jona-Lasinio (NJL) model~\cite{Nambu.1961}, which describes the spontaneous generation of mass for quarks via the formation of a chiral condensate.
In a Weyl semimetal, this model can be used to describe the transition into a charge-density wave phase, dubbed an axionic insulator~\cite{Nenno.2020}.
As in the NJL model, this similarly gaps out fermionic quasiparticles and produces a collective soft Goldstone mode (analogous to the pion in particle physics). 
Due to the chiral anomaly, these Goldstone modes become endowed with a dynamic response resembling a fluctuating ``$\theta$-term" in analogy to high-energy physics~\cite{Li.2010, Wan.2011,Wang.2013,Liu.2013,Roy.2015,Laubach.2016,Wieder.2020,Sehayek.2020,Yu.2021,McKay.2021,Zeng.2022,Sekine.2016a,Sekine.2016b,Sekine.2021,Tserkovnyak.2021}. 
Such a response was even reported experimentally~\cite{Gooth.2019,Mu.2021} in the compound (TaSe$_4$)$_2$I, which has both a charge-density wave and Weyl fermions~\cite{Tournier-Colletta.2013,Shi.2019}, although this remains heavily debated~\cite{Cohn.2020,Sinchenko.2022}.

While the mean-field phase diagram of the NJL model is known well at zero temperature and density, it is still uncertain what the fate of this system is at finite temperature, finite density, and beyond mean-field~\cite{Kleinert.2000,Ripka.2000}.
In particular, it is believed that under certain conditions the NJL model may exhibit inhomogeneous chiral condensate order~\cite{Nickel.2009,Pisarski.2021}.
In condensed matter physics, the corresponding parameter regimes are much milder in comparison to their high-energy analogues \textemdash temperature can easily be tuned to the scale of ``meson disassociation" (corresponding simply to melting the charge-density wave), and finite particle density is achievable through combinations of doping and magnetic field interaction~\cite{Tserkovnyak.2021}.
Therefore, examining the dynamical axion response in such an inhomogeneous chiral condensate is readily within reach.

In this {\it Letter}, we study a minimal model for a correlated Weyl semimetal and show that it exhibits a finite-momentum instability under simple conditions.
We then derive an effective long-wavelength model for the resulting finite-momentum condensate and propose a characteristic optical signature in light-scattering due to the fluctuating axionic response.
In particular, we study a model of two isotropic Weyl points interacting with a mean-field chiral condensate in the presence of both a chiral chemical potential $\kappa$ as well as a regular chemical potential $\mu$.
We show that in certain temperature and density regimes this model maps on to the Fulde-Ferrell-Larkin-Ovchinikov (FFLO) phases of superconductivity in a large Zeeman field, which is known to result in finite-momentum condensation~\cite{Fulde.1964}.
We then propose an appropriate Lifschitz model to describe the axionic response of such a strongly-fluctuating phase and predict characteristic signatures in inelastic light scattering. 

{\it Model}\textemdash
We consider an effective model for Weyl fermions, with operator $\Psi(x) = (\Psi_R(x),\Psi_L(x))^T$, interacting with a mean-field chiral condensate ${\Delta \sim \langle \Psi_{R}^\dagger \Psi_L\rangle }$~\cite{Liu.2013,Wang.2013,Sehayek.2020,McKay.2021,Nambu.1961}.
In the presence of both a chiral chemical potential $\kappa$ and a regular chemical potential $\mu$ the effective Matsubara Lagrangian is 
\begin{widetext}
\begin{equation}
\label{eqn:Weyl}
    \mathcal{L} = \overline{\Psi}\left[  \frac{\partial}{\partial \tau}-\mu +\tau_3 {\bm \sigma}\cdot\left(-i\nabla- \tau_3 \mathbf{Q}/2 \right)  - \kappa \tau_3 + \overline{\Delta}e^{-i\mathbf{Q}\cdot\mathbf{r}}\tau^+ + \Delta e^{i\mathbf{Q}\cdot\mathbf{r}} \tau^-\right] \Psi + \frac{|\Delta|^2}{g}.
\end{equation}
\end{widetext}
This model features two Weyl points located at $\pm \mathbf{Q}/2$ in reciprocal space (we take the right-handed fermions to reside at $+\mathbf{Q}/2$), separated by $2\kappa$ in energy.
This is illustrated in Fig.~\ref{fig:main-figure}(a),(b) for the cases of $\mu = 0$ (compensated pockets), and $\mu >0$ (uncompensated pockets). 
Each Weyl spinor carries spin 1/2 and is characterized by the Pauli matrices $\bm \sigma$, while we reserve Pauli matrices $\bm \tau$ for the chirality quantum number. 
For simplicity, we take the dispersion near each Weyl cone to be isotropic with Fermi velocity $v_F = 1$.

We note that time-reversal symmetry acts in this model as $\mathcal{T} = \tau_1 i\sigma_2$, along with $\mathbf{p}\to- \mathbf{p}$ and the usual complex conjugation, whereas inversion symmetry acts as $\mathcal{I} = \tau_1$ and $\mathbf{p}\to- \mathbf{p}$.
Therefore having only two Weyl points requires that the Hamiltonian explicitly breaks time-reversal symmetry, while the presence of the finite chiral chemical potential requires the additional breaking of inversion symmetry.
In the following, we perform a chiral gauge transformation to remove the fast-varying component $e^{i\mathbf{Q}\cdot\mathbf{r}}$, absorbing it into the spinors $\Psi$, and obtaining a resulting theory for only the slowly-varying envelope, as originally done in Ref.~\cite{Wang.2013}.

{\it Mean-Field}\textemdash
The mean-field solution to this model follows analogously to the case of Bardeen-Cooper-Schrieffer (BCS) superconductivity; we first integrate out the Weyl fermions and take the saddle point of the resulting action for the order parameter $\Delta$.
Crucially, at finite chiral chemical potential $\kappa$ there is a finite density of states at the Fermi level $\nu(E_F) = \frac{\kappa^2}{2\pi^2}$ that induces a weak-coupling instability~\cite{Braguta.2019}.
This is diagnosed by solving the gap equation   
\begin{multline}
    \frac{\Delta}{g} = -\int_p \tr \tau \mathbb{G}(p) \\
    = -T\sum_{i\epsilon_m}\int_{\bf p}\sum_{\sigma = \pm 1} \frac{\Delta}{(i\epsilon_m+\mu)^2 - (\sigma |\mathbf{p}| - \kappa )^2 - |\Delta|^2},
\end{multline}
where $\mathbb{G}(p) = \left(i\epsilon_m +\mu - ( {\bm \sigma}\cdot\mathbf{p} - \kappa )\tau_3  - \Delta \tau^+ - \overline{\Delta}\tau^- \right)^{-1}$ is the mean-field electronic Green's function in terms of the fermionic Matsubara frequency $\epsilon_m = 2\pi T (m+\frac12)$.

When $\mu = 0$ the origin of the weak-coupling instability is clear: it arises from the gapless positive-helicity states, which have a normal-state propagator of $(i\epsilon_m - \xi_{\bf p} )^{-1}$, with $\xi_{\bf p} =  |\mathbf{p}|-\kappa$.
As $\xi_{\bf p} \to 0$ upon approaching the Fermi surface, this generates the famous ``Cooper logarithm" in the pairing susceptibility
which for sufficiently low temperature will always diverge, guaranteeing a condensate will develop. 
The presence of the other helicity states introduces additional corrections that are suppressed when the chiral chemical potential $\kappa$ is much larger than the cutoff on the BCS-type interaction (in superconductors this is usually the Debye frequency).
In this limit the pairing interaction can be safely projected on to the Fermi surface and the minority helicity bands can be projected out. 
We also comment here that, while the bare Weyl fermions with $\kappa= \mu = 0$ exhibit an effective Lorentz invariant dispersion, the presence of $\kappa$ or $\mu$ produces a finite Fermi surface which explicitly breaks this Lorentz invariance by selecting a preferential frame of reference.

We now restore the finite chemical potential $\mu$ and note that the analogous term in the BCS superconductor (which corresponds to a finite Zeeman field) is known to quench spin-singlet superconductivity and ultimately suppress condensation. 
In our case, the role of the chemical potential is to imbalance the two Fermi surfaces, leading to a difference of $2\mu$ in their Fermi momenta. 
While for sufficiently large imbalances the homogeneous condensate is suppressed, the chiral pairing will in fact persist at finite momentum.
This is caused by the partial nesting of the Fermi surface which typically forms a condensate of either a standing-wave, known as the Larkin-Ovchinikov (LO) phase in space, or a plane-wave, known as the Fulde-Ferrell (FF) phase; such a mechanism was originally found to occur in superconductors and is by now a well understood phenomenon~\cite{Fulde.1964,Kinnunen.2018,Casalbuoni.2004,Radzihovsky.2011,Lake.2021}.

The instability towards finite momentum can be diagnosed at the level of a Ginzburg-Landau theory for the chiral condensate as a function of chemical potential~\cite{Buzdin.1997,Radzihovsky.2011,Combescot.2002,Casalbuoni.2004,Kinnunen.2018,Radzihovsky.2009}. 
The resulting free energy ${\mathcal{F} = \sum_{\bf q} \mathscr{L}(\mathbf{q})|\Delta_{\bf q}|^2 }$ is determined by the leading Gaussian, momentum-dependent susceptibility~\cite{Buzdin.1997,Radzihovsky.2009}, given in the random phase approximation by (see Supplemental Material for more detail)
\begin{equation}
    \mathscr{L}(\mathbf{q}) = \frac{1}{g} - \int_p \tr \mathbb{G}_0(p)\tau^+ \mathbb{G}_0(p+q) \tau^-
\end{equation}
The result of this calculation is shown for two different values of $\mu$ in Fig.~\ref{fig:main-figure}(c) and (d).
For sufficiently small $\mu$, the system will still condense at zero momentum, though the critical temperature $T_c$ drops with increasing $\mu$.
On the other hand, for sufficiently large $\mu$ the condensate at $\mathbf{q} =0$ is suppressed and, instead, a finite-momentum condensate develops at a critical temperature $T_c$ and momentum $q_\star$ set by $\mu$. 

{\it Fluctuating Axion Response} \textemdash In general, the phase diagram of such an inhomogeneous phase is quite complex~\cite{Buzdin.1997,Casalbuoni.2004,Radzihovsky.2009,Radzihovsky.2011,Combescot.2002,Kinnunen.2018}.
Here we consider a phenomenological analysis of this phase and assume a plane-wave solution, i.e., the FF ground-state~\cite{Fulde.1964}, though the generalization to more complex liquid-crystalline condensates would be interesting. 
Keeping the amplitude $|\Delta|$ fixed, the minimal model describing the long-wavelength dynamics of the soft phase $\partial_\mu\theta(x)$ in the  finite-momentum condensate requires terms up to fourth order in the spatial derivatives.
Generically, this is given by
\begin{multline}
    \label{eqn:ff-nlsm}
\mathcal{L} = \frac12 K (\partial_\tau \theta)^2 + \frac{J}{4}( (\nabla \theta)^2 - q_\star^2)^2 + \frac{J'}{2}(\nabla^2 \theta)^2 \\
+ i\frac{e^2}{8\pi^2} \left(\theta+\mathbf{Q}\cdot\mathbf{r} \right)\mathbf{E}\cdot\mathbf{B}
\end{multline}
where $K$ describes the chiral compressibility, $J$ and $J'$ describe the compressional and bending moduli of the condensate, respectively (in analog with the theory of liquid crystals~\cite{Radzihovsky.2009}), $q_\star$ is determined by the minimum of the free energy~\footnote{We comment here that we have considered for simplicity the collective modes in the Fulde-Ferrell phase, which differs slightly from the Larkin-Ovchinikov model considered in Ref.~\cite{Radzihovsky.2009}.}, while $\mathbf{E} =\partial_\tau\mathbf{A} + \nabla A_0 $ and $\mathbf{B} = \nabla \times \mathbf{A}$ are the electric and magnetic fields in terms of the gauge potentials.
In general, $K$, $J$, and $J'$ are functions of the microscopic parameters and the condensate amplitude $|\Delta|$, though, dimensional analysis suggests $K \sim \nu(E_F) \Delta^2 $ and $J/K,J'/K \sim v_F^2/q_\star^2$.

The last term in Eq.~\eqref{eqn:ff-nlsm} is due to the chiral anomaly and results from the chiral gauge transformation in the presence of gauge fields $\mathbf{E}\cdot\mathbf{B}$ (see Supplemental Material).
This has contributions from: (i) the fluctuating phase $\theta$, and (ii) the static part of the band structure $\mathbf{Q}\cdot\mathbf{r}$. 
While the latter yields a homogeneous anomalous Hall effect (AHE), here we are more interested in the phase mode $\theta$, which is expected to fluctuate very strongly.
Due to the chiral anomaly, these fluctuations also couple to the electromagnetic field and, in particular, contribute to the AHE.

To diagnose the effect of these fluctuations we perform a Gaussian approximation where the ground state is obtained by $\theta =  z q_\star$~\footnote{In the future, a more sophisticated renormalization group procedure should probably be employed due to the highly nonlinear, strongly fluctuating nature of the model~\eqref{eqn:ff-nlsm}.
In particular, topological defects are expected to play an important role~\cite{Radzihovsky.2009}, which we will neglect for the moment.} (without loss of generality, we choose $\mathbf{e}_z$ to put the momentum gradient on).
In this case, the system has spontaneously broken rotational symmetry, although in a more realistic treatment rotational symmetry would be broken by the crystal, e.g. to reduced uniaxial or planar symemtry, which may impact the ordering of the phase.
The finite additional phase gradient due to $q_\star$ offsets the effective momentum-space separation $\mathbf{Q}\to\mathbf{Q} + q_\star \mathbf{e}_z$ to a slightly different nesting vector. 
This change can be detected in magnetoelectric transport where by driving the system into the finite-momentum condensate shifts the value of the Hall coefficient.

Finally, we turn our attention to the fluctuating part $\theta(x) = q_\star z + \delta \theta_q e^{i\mathbf{q}\cdot\mathbf{r} - i \Omega(\mathbf{q}) t}$, and linearize in $\delta \theta_q$.
The phase fluctuations exhibit a dispersion relation which is highly-anisotropic, with
\begin{equation}
\Omega(\mathbf{q}) = \sqrt{ v^2 q_z^2 +v'^2 |\mathbf{q}|^4 }.
\label{eq:dispersion}
\end{equation}
where $v = \sqrt{2 J q_\star^2/K}$ is the longitudinal sound velocity and $v' = \sqrt{J'/K}$ is the higher-order transverse velocity.
While longitudinal oscillations with $\mathbf{q} \parallel\mathbf{e}_z$ disperse linearly due to the compressional modulus $J$, the transverse modes with $\mathbf{q}\perp\mathbf{e}_z$ (which lead to fluctuations of the orientation of the phase gradient) are very soft, dispersing quadratically with the bending modulus $J'$. 
As a result we expect a very large number of long-wavelength fluctuations which will reduce the long-range correlations in the anomalous Hall response. 

{\it Optical Detection}\textemdash
We now explore the fluctuating part of the Hall response and, in particular, determine how these fluctuations can produce signatures in light scattering experiments.
Specifically, we consider Brillouin light scattering, which describes the scattering of light off of acoustic modes and involves an exchange of both energy and momentum. 
In the presence of a fluctuating axion phase $\delta \theta$ the modified current in Ampere's Law is given by
\begin{equation}
    \mathbf{J} = - \frac{e^2}{4 \pi^2} (\partial_t\delta \theta) \mathbf{B} - \frac{e^2}{4 \pi^2 } (\nabla \delta \theta)\times \mathbf{E},
\end{equation}
where we disregarded the mean-value of $\theta$ for simplicity.
These contributions can be measured using Dynamic Light Scattering~\cite{Berne.1976}, which is sensitive to the fluctuations of the dielectric constant, and hence can be used to measure the fluctuations of the Hall conductivity.

The linearized fluctuations $\delta \theta$ obey the highly-anisotropic dispersion relation $\Omega({\bf q})$, see Eq.~\eqref{eq:dispersion}, where $v$ and $v'$ are generally expected to be of order of the Fermi velocity $v_F$ of the underlying Weyl cones~\footnote{This may be more complex in principle, with a possible dependence on the chemical potential $\mu$ that can lead to a significant slowing of the velocities near the critical point where the normal condensate phase gives way to the FF phase. We disregard this subtlety here.}.
As such, the excitation frequency can be taken to be much smaller than the optical probing frequency, hence, to lowest order in $v_F/c$ the scattering is elastic.
In this case contributions from $\partial_t \delta \theta$ vanish and scattering of light is dominated by the fluctuating Hall conductivity 
\begin{equation}
    \delta \sigma_{ab}(\mathbf{q}) = \frac{e^2}{4\pi^2}\delta\mathbf{n}_c(\mathbf{q}) \epsilon_{abc} ,
\end{equation}
where $\epsilon_{abc}$ is the three-dimensional Levi-Civita tensor and $\delta \mathbf{n}= \nabla \delta \theta$.
In this limit, the light scattering is sensitive to the static structure factor, calculated in linear response as 
\begin{equation}
    \langle \delta \sigma_{jk}(\mathbf{q}) \delta \sigma_{lm}(-\mathbf{q}) \rangle = T \left(\frac{e^2}{4\pi^2}\right)^2 \epsilon_{jka}\epsilon_{lmb}\frac{ q_a q_b}{ 2 J q_\star^2 q_z^2 + J' \mathbf{q}^4 }.
\end{equation}
In addition to the polarization dependence due to the Levi-Civita symbols, the static structure factor has a characteristic pinch-point singularity as $\mathbf{q} \to 0$, whereby it diverges differently depending on the angle of $\mathbf{q}$ relative to the nematic axis $\mathbf{\hat{e}}_z$.

{\em Conclusion}\textemdash
We have explicitly verified that in a simple Weyl semimetal with both broken $\mathcal{T}$ and $\mathcal{P}$ symmetries, where an imbalance in the carrier density of the two chiralities can naturally arise, an instability of the chiral condensate appears towards finite momentum (similar to the Fulde-Ferrell-Larkin-Ovchinikov phase in a spin-polarized Fermi superfluid).
Due to the chiral anomaly, the spatial fluctuations of this phase lead to a characteristic fluctuating Hall conductivity that gives rise to characteristic signatures in light-scattering experiments. 

In the future, it will be important to consider a more detailed microscopic model in order to make contact with experiments, as well as treat the non-Gaussian nature of the fluctuations more quantitatively.
In particular, studying the role of anisostropy, multiple Weyl points, electron-electron and electron-phonon interactions, and disorder will be important for obtaining a more precise phase diagram and prediction for light scattering. 
Our results show that solid-state systems are ripe for studying the interplay of strong spatial fluctuations and inhomogeneity, topology, and condensation~\cite{Lake.2021}.
This may offer insight into the complex phase diagrams in nuclear and particle physics at finite temperature and density, or unveil new unconventional ``topological" electronic liquid-crystal phases.

\begin{acknowledgements}
The authors would like to acknowledge fruitful discussions with Alireza Parhizkar, Dennis Nenno, Andrey Grankin, B. Andrei Bernevig, Claudia Felser, Olivia Liebman, Kenneth Burch, Tibor Rakovszky, Eugene Demler, Rahul Nandkishore, Johannes Gooth, Nick Poniatowski, and Justin Wilson.
This work is supported by the Quantum Science Center (QSC), a National Quantum Information Science Research Center of the U.S. Department of Energy (DOE). 
I.P. was supported by the Early Postdoc mobility grant from the Swiss National Science Foundation (SNSF) under project ID P2EZP2\_199848.
\end{acknowledgements}

\bibliography{references,references-2}

\appendix

%%%%%%%%%%%%%%%%%%%%%%%%%%%%%%%%%%%%%%%%%%%%%%%%%%%%%%%%%%%%
%%%%%%%%%%%%%%%%%%%%%%%%%%%%%%%%%%%%%%%%%%%%%%%%%%%%%%%%%%%%
%%%%%%%%%%%%%%%%%%%%%%%%%%%%%%%%%%%%%%%%%%%%%%%%%%%%%%%%%%%%
%%%%%%%%%%%%%%%%%%%%%%%%%%%%%%%%%%%%%%%%%%%%%%%%%%%%%%%%%%%%
\section{Mapping to Weyl Equation}
\label{app:weyl}
Here we outline the mapping from the Euclidean path integral and Hamiltonian prescription to the more conventional relativistic form for the Weyl equation. 
We do this for the long-wavelength field $\Psi$ which we get after gauging away the momentum space separation.
We also take $v_F = 1$.

\begin{equation}
    \mathcal{L} = \overline{\Psi}\left[ D_0 -i \tau_3 {\bm \sigma}\cdot \mathbf{D} + \Delta \tau^+ + \overline{\Delta}\tau^- \right] \Psi.
\end{equation}
We now define the Dirac spinors 
\begin{subequations}
\begin{align}
    & \psi = \Psi =  \begin{pmatrix} \psi_R \\
    \psi_L \\
    \end{pmatrix}  \\
    & \overline{\psi} = \overline{\Psi}(i\tau_2) = \left( \overline{\psi}_L , -\overline{\psi}_R \right). 
\end{align}
\end{subequations}
We now see that the action becomes (explicitly writing out the gauge field)
\begin{equation}
    \mathcal{L} = \overline{\psi}(-i\tau_2)\left[\partial_\tau - i eA_0  + \tau_3 {\bm \sigma}\cdot(-i\nabla - e\mathbf{A}) + \Delta \tau^+ + \overline{\Delta}\tau^- \right] \psi.
\end{equation}
We now introduce the four $\gamma^\mu$ matrices as 
\begin{subequations}
\begin{align}
    & \gamma^0 = \tau_2 \\
    & \gamma^j = \tau_1 \sigma^j ,
\end{align}
\end{subequations}
and $\gamma^5 = \gamma^0 \gamma^1 \gamma^2 \gamma^3 = \tau_3 $.
Since we are working in Euclidean spacetime we have 
\begin{equation}
    \{\gamma^\mu,\gamma^\nu\} = 2\delta_{\mu\nu},
\end{equation}
which is the metric with standard Euclidean signature, and $(\gamma^\mu)^\dagger = \gamma^\mu$.

We thus obtain the usual Dirac equation with chiral gauge field coupled to the mass as 
\begin{equation}
    \mathcal{L} = \overline{\psi}\left[\gamma^0 (-i\partial_\tau - eA_0 ) + {\bm \gamma}\cdot(-i\nabla - e\mathbf{A}) -|\Delta|\gamma^5  e^{i\theta(x) \gamma^5 }\right] \psi.
\end{equation}
We define the Dirac slashed operator 
\begin{equation}
    \slashed{D} = \gamma^\mu ( \partial_\mu - ie A_\mu ), 
\end{equation}
such that 
\begin{equation}
    \mathcal{L} = \overline{\psi}\left[ -i \slashed{D} - |\Delta| \gamma^5 e^{i\gamma^5 \theta(x)} \right]\psi.
\end{equation}

For future reference we have
\begin{equation}
    \{\gamma^5,\slashed{D}\} =0.
\end{equation}
Finally, we note that in the absence of electromagnetic gauge field we have 
\begin{equation}
    \slashed{D}^2 = \slashed{\partial}^2 = \left[ (-i\epsilon_m)^2 + (i\mathbf{q})^2 \right] .
\end{equation}
Upon continuation to real time we would then have 
\begin{equation}
   \left[ (-i\epsilon_m)^2 + (i\mathbf{q})^2 \right] \to \varepsilon^2 - \mathbf{q}^2, 
\end{equation}
the appropriate Lorentz invariant dispersion relation. 

\section{Anomaly via Fujikawa Method}
\label{app:fujikawa}
We consider the attempted chiral gauge transformation and its effect on the integration measure.
We follow Ref.~\cite{Fujikawa.2004}, and in particular focus on the approach based on evaluation of the Jacobian. 
To this end, we consider the transformation of the integration measure $\mathcal{D}[\psi,\overline{\psi}]$ under the transformation 
\begin{subequations}
\begin{align}
    & \psi(x) = e^{i \alpha(x) \gamma_5 } \eta(x) \\
    & \overline{\psi}(x) = \overline{\eta}(x) e^{i \alpha(x) \gamma_5 } .
\end{align}
\end{subequations}
To evaluate the Jacobian, we must regularize our integration measure, which we write in terms of Grassman valued normal modes.
Let us use as a set of basis functions the gauge-invariant eigenspectrum of the operator $-i\slashed{D}$, with 
\begin{equation}
    -i\slashed{D} \phi_n (x) = \lambda_n \phi_n(x)
\end{equation}
and the corresponding fermionic field operator is 
\begin{equation}
    \psi(x) = \sum_n \psi_n \phi_n(x).
\end{equation}
We then may define the gauge invariant functional integration measure as 
\begin{equation}
    \mathcal{D}[\psi,\overline{\psi}] = \prod_n d \psi_n d\bar{\psi}_n .
\end{equation}
Note that $\{ \gamma_5, \slashed{D}\} = 0$.

The chiral gauge transformation acts to change the normal mode eigenbasis.
We may write it in terms of the same basis functions as 
\begin{equation}
    \eta(x) = \sum_n \eta_n \phi_n(x) .
\end{equation}
The gauge transformation induces a linear transformation on the Grassman coefficients such that 
\begin{equation}
    \psi_m = \underbrace{\int d^4 x \overline{\phi}_m(x) e^{i\alpha(x) \gamma_5 } \phi_n(x)}_{\Xi_{mn}[\alpha]} \eta_n .
\end{equation}
Due to standard Grassman integration rules, the transformation on the integration measure induced by this change of variables is the inverse of the determinant, such that (recall both fields transform in the same way)
\begin{equation}
    \mathcal{D}[\eta,\overline{\eta}] = \mathcal{D}[\psi,\overline{\psi}] \left( \Det\Xi \right)^2.
\end{equation}
We evaluate the determinant as 
\begin{equation}
    \Det \Xi = \exp \Tr \log \Xi.
\end{equation}
This must be regularized in order to be evaluated.
The most straightforward way is via the heat-kernel method, where we impose a cutoff on modes which have a large eigenvalue of the {\em gauge invariant} derivative. 
We have 
\begin{equation}
\Tr \log \Xi = \sum_n  e^{-\lambda_n^2/\Lambda^2} \left(\log \Xi\right)_{nn} .
\end{equation}
The gauge transformation is commuting in the local eigenbasis so we can evaluating the matrix logarithm in real-space.
This gives 
\begin{multline}
\Tr \log \Xi = \sum_n  e^{-\lambda_n^2/\Lambda^2}\int d^4x \overline{\phi}_m(x) i \alpha(x)\gamma_5 \phi_n(x) \\
\sim i \int d^4 x \alpha(x) \sum_n e^{-\lambda_n^2/\Lambda^2}  \overline{\phi}_n(x) \gamma_5 \phi_n(x) .
\end{multline}
This is expressed as the coupling between the phase and the anomalous action via 
\begin{equation}
    \Tr \log \Xi = i \int d^4 x \alpha(x) \mathcal{A}(x)
\end{equation}
with 
\begin{equation}
    \mathcal{A}(x) =\lim_{x'\to x} \sum_n \overline{\phi}_n(x) \gamma_5 e^{\slashed{D}^2/\Lambda^2} \phi_n(x') = \tr \gamma_5 \left(e^{\slashed{D}^2/\Lambda^2}\right)(x,x') .  
\end{equation}

We use
\begin{equation}
\slashed{D}^2 = \gamma^\mu \gamma^\nu \left( \partial_\mu - i e A_\mu \right)\left( \partial_\nu - ie A_\nu\right) = D_\mu D^\mu - \frac{ie}{2}\gamma^\mu \gamma^\nu F_{\mu\nu} .
\end{equation}
We further have $D_\mu D^\mu = \partial^2 -i e \{ \partial^\mu ,A_\mu \} - e^2 A^2 $, which yields the gauge invariant spectrum of an equivalent charged boson. 

We then find, using the standard BCH formula
\begin{widetext}
\begin{equation}
    \mathcal{A}(x) = \sum_n \overline{\phi}_n(x) \gamma_5 e^{\slashed{D}^2/\Lambda^2} \phi_n(x) = \tr \gamma_5 e^{D^2/\Lambda^2}\left(1 + \frac{ - \frac{ie}{2}\gamma^\mu \gamma^\nu F_{\mu\nu}}{\Lambda^2} + \frac12 \frac{\left( - \frac{ie}{2}\gamma^\mu \gamma^\nu F_{\mu\nu}\right)^2}{\Lambda^4} - \frac{ie}{4\Lambda^4}\gamma^\mu\gamma^\nu[ D^2, F_{\mu\nu}]\right)  
\end{equation}
\end{widetext}
The only term which has a non-zero trace against the chiral gamma matrix is the middle term with $F^2$. 
Also note that this is still composed only of manifestly gauge-invariant terms. 
We have 
\begin{equation}
    \mathcal{A}(x) = \tr \gamma_5 e^{D^2/\Lambda^2} \frac12 \frac{\left( - \frac{ie}{2}\gamma^\mu \gamma^\nu F_{\mu\nu}\right)^2}{\Lambda^4}
\end{equation}
This simplifies to 
\begin{equation}
    \mathcal{A}(x) = -\frac{e^2}{8}\tr\left(\gamma_5\gamma^\mu \gamma^\nu\gamma^\alpha\gamma^\beta\right) F_{\mu\nu}F_{\alpha\beta} \left[ \frac{e^{D^2/\Lambda^2} }{\Lambda^4} \right].  
\end{equation}
Here we have used the fact that for slowly-varying field configurations, the field-strength tensor can be treated as a constant, leaving only the heat-kernel itself as the remaining object to be evaluated. 
We also have $ \tr\left(\gamma_5\gamma^\mu \gamma^\nu\gamma^\alpha\gamma^\beta\right) = 4 \epsilon^{\mu\nu\alpha\beta}$, so that we have
\begin{equation}
    \mathcal{A}(x) = -\frac{e^2}{2}\epsilon^{\mu\nu\alpha\beta} F_{\mu\nu}F_{\alpha\beta} \left[ \frac{e^{D^2/\Lambda^2} }{\Lambda^4} \right].  
\end{equation}
Now, to evaluate the divergent term we expand in a plane-wave basis. 
It can be seen that the corrections arising from the gauge field in the plane-wave basis are of order $1/\Lambda$ and thus vanish in the long-wavelength limit so that we can evaluate using $D^2 = -p^2$ so that 
\begin{equation}
\frac{e^{D^2/\Lambda^2} }{\Lambda^4}(x,x) = \frac{1}{\Lambda^4}\int_p e^{-p^2/\Lambda^2} + O(1/\Lambda^5) = \frac{1}{16\pi^2} +O(1/\Lambda) .  
\end{equation}

We then obtain 
\begin{equation}
    \mathcal{A}(x) = -\frac{e^2}{32 \pi^2} \epsilon^{\alpha\beta\mu\nu} F_{\alpha\beta}F_{\mu\nu}. 
\end{equation}
Finally, we note that taking both copies in to account, this result is the additional contribution to he effective action (strictly valid at zero temperature) of 
\begin{equation}
    \mathcal{S}_{\rm anomaly} = + i \frac{e^2}{16 \pi^2} \int d^4x \alpha(x) F_{\alpha\beta}F_{\mu\nu} \epsilon^{\alpha\beta\mu\nu}.
\end{equation}
We express this in terms of the electric and magnetic fields (recall the scalar potential enters with opposite sign relative to usual relativistic convention) as 
\begin{equation}
    F_{\alpha\beta}F_{\mu\nu} \epsilon^{\alpha\beta\mu\nu} = 4(\partial_\tau \mathbf{A} + \nabla A_0 )\cdot \nabla \times \mathbf{A}.
\end{equation}
Thus, we get imaginary time axion term of 
\begin{equation}
     \mathcal{S}_{\rm axion} = + i \frac{e^2}{4 \pi^2} \int d^4x \alpha(x) \mathbf{B}\cdot(\partial_\tau\mathbf{A} + \nabla A_0).
\end{equation}
Note to return to the real-time result, we take $d\tau = +id t , \partial_\tau = -i \partial_t, A_0 = - i\phi$ so that we see this term becomes 
\[
     e^{-S_{\rm axion}} = e^{- i \frac{e^2}{4 \pi^2} \int d^4x \alpha(x) \mathbf{B}\cdot(-\partial_t\mathbf{A} - \nabla \phi)},
\]
so that we may identify the real time action in terms of electric and magnetic fields as 
\begin{equation}
    S_{\rm axion} = - \frac{e^2}{4\pi^2} \int d^4 x \alpha(x)\mathbf{B}\cdot\mathbf{E}. 
\end{equation}
Finally, we must connect the infinitesimal gauge transformation with the complete transformation needed to remove the axion phase from the mass terms. 
We can see that performing an infinitesimal transformation is going to be additive in this case, since it is linear in the transformation $\alpha(x)$.
Therefore, we can simply replace $\alpha(x) = \frac12 (\theta(x) + \mathbf{Q}\cdot\mathbf{r})$ in the above to obtain the final result
\begin{equation}
     \mathcal{S}_{\rm axion} = + i \frac{e^2}{4 \pi^2} \int d^4x \frac12 (\theta(x) + \mathbf{Q}\cdot\mathbf{r}) \mathbf{B}\cdot(\partial_\tau\mathbf{A} + \nabla A_0).
\end{equation}

\section{Chiral Condensate Cooperon}
Here we elaborate slightly on the calculation of the chiral condensate ``Cooperon" (i.e. collective-mode fluctuation propagator) in the presence of the chiral and regular chemical potentials. 
We expand the NJL action up to quadratic order in the order parameter $\Delta(q)$ in the Random Phase Approximation (RPA) to obtain 
\begin{equation}
    \mathcal{S} = \sum_q \mathscr{L}(q) |\Delta_q|^2. 
\end{equation}
We find the standard expression
\begin{equation}
    \mathscr{L}(q) = \frac{1}{g} + \tr \int_p \tau^+ \mathbb{G}_0(p+q) \tau^- \mathbb{G}_0(p) .
\end{equation}
where
\begin{equation}
    \mathbb{G}_0(p) =\left[ i \epsilon_m + \mu - \tau_3 (\mathbf{p}\cdot{\bm \sigma} -\kappa)\right]^{-1}
\end{equation}
is the normal state Weyl-fermion propagator. 
We focus on the momentum dependence of this, to wit we set $\omega_m = 0$ in the Cooperon and compute
\begin{widetext}
\begin{equation}
    \mathscr{L}(\mathbf{q},\omega = 0) = \frac{1}{g} - \int_{p} \tr \left[  \left(i\epsilon_m + \mu - ({\bm \sigma}\cdot( \mathbf{p}+\mathbf{q})-\kappa) \right)^{-1}  \left(i\epsilon_m + \mu + ({\bm \sigma}\cdot( \mathbf{p})-\kappa) \right)^{-1} \right].
\end{equation}
\end{widetext}
For small $\mathbf{q}$ this can be diagonalized in terms of the helicities $\sigma = {\bm \sigma}\cdot\mathbf{p}/|\mathbf{p}| = \pm 1$; there are two for each chiral Fermi pocket.

On the one hand, for the positive helicity $\sigma = +1$ the electron propagator will exhibit a resonance upon approaching the Fermi surface, whereupon the dispersion $|\mathbf{p}| -\kappa $ crosses through zero. 
This will therefore produce a strong contribution to the collective dynamics, yielding the Cooper logarithm in the absence of $\mu$. 
The other helicity $\sigma = -1$ will always remain gapped and buried below the Fermi surface, with a quasiparticle excitation energy at least of order $\kappa$. 

On the other hand, the BCS-like interaction, which is characterized by attraction $g$ is not valid throughout the entirety of momentum space, but rather result from projecting of a more microscopically accurate interaction on to the Fermi surface.
This is only valid for momenta near the Fermi surface, implying the integral on $\mathbf{p}$ should be cutoff at $\Lambda$.
In the case where $\Lambda \gtrsim \kappa$ we find that in general both helicities participate in the interaction and the system is more complicated.
We focus on the simpler case where $\Lambda \ll \kappa$, in which case only states at the Fermi surface participate. 
We can therefore discard the $\sigma = -1$ helicity and project onto the $\sigma = +1$ states, which have ${\bm \sigma} \to \hat{\mathbf{p}}=\mathbf{p}/|\mathbf{p}|$.
This yields 
\begin{equation}
    \mathscr{L}(\mathbf{q},\omega = 0) = \frac{1}{g} + \int_{p}  \frac{1}{(i\epsilon_m + \mu + \hat{\mathbf{p}}\cdot\mathbf{q})^2 - (|\mathbf{p}|-\kappa)^2 }.
\end{equation}
This can be evaluated in the usual quasiclassical approximation by writing $\mathbf{q}\cdot \hat{\mathbf{p}} = q u$ where $u = \cos \theta$ is the scattering cosine, and $\xi = |\mathbf{p}|-\kappa$, producing 
\begin{equation}
    \mathscr{L}(\mathbf{q},\omega = 0) = \frac{1}{g} + T \sum_{i\epsilon_m} \nu(E_F) \int d\xi \int_{-1}^{1}\frac{du}{2}  \frac{1}{(i\epsilon_m + \mu + q u)^2 - \xi^2 }.
\end{equation}
Here we have the density-of-states at the Fermi level $\nu(E_F) = \kappa^2/(2\pi^2)$ enter.
We can remove the dependence on $g$ by renormalizing the scattering length with reference to the transition temperature at $\mu = 0,q = 0$, defined by the equation 
\begin{equation}
   \nu(E_F)\log(T/T_c^{(0)}) = \frac{1}{g} + T \sum_{i\epsilon_m} \nu_F \int d\xi \frac{1}{(i\epsilon_m)^2 - \xi^2 }.
\end{equation}
with usual BCS result for $T_c^{(0)}\sim \Lambda e^{-1/(g \nu(E_F))}$.
We then find the UV convergent expression of 
\begin{widetext}
\begin{equation}
    \mathscr{L}(\mathbf{q},\omega = 0) = \nu(E_F) \left[ \log(T/T_c^{(0)}) + T \sum_{i\epsilon_m} \int d\xi \int_{-1}^{1}\frac{du}{2}\left(  \frac{1}{(i\epsilon_m + \mu + q u)^2 - \xi^2 } - \frac{1}{(i\epsilon_m)^2 - \xi^2} \right)\right].
\end{equation}
Utilizing 
\begin{equation}
    \int d\xi \frac{1}{\xi^2 + (\epsilon_m + iz)^2 } = \frac{\pi \textrm{sign}(\epsilon_m)}{\epsilon_m + iz}
\end{equation}
and simplifying we obtain the result 
\begin{equation}
    \mathscr{L}(\mathbf{q},\omega = 0) = \nu(E_F) \left[ \log(T/T_c^{(0)}) - 2\pi T \sum_{\epsilon_m>0} \int_{-1}^{1}\frac{du}{2}\left(  \frac{\epsilon_m}{\epsilon_m^2 + (\mu + q u)^2 } - \frac{1}{|\epsilon_m|} \right)\right].
\end{equation}
This is easily evaluated as a function of $\mathbf{q}$ by numerically summing Matsubara frequencies and performing the solid-angle integral on $u$ numerically, since the integral is both IR and UV convergent. 
\end{widetext}
We find for sufficiently large $\mu$ that the transition (signified by $\mathscr{L} = 0$) occurs at a finite momentum $|\mathbf{q}|=q_\star$.
The exact form of $\mathscr{L}(\mathbf{q})$ near $q_\star$ is not particularly clear, and furthermore would require numerical evaluation in a real system.
We can generically conclude however that as $\mu$ passes through the critical $\mu_c$, $q_\star$ would soften at $\mu_c$ and then smoothly rise again, implying that the characteristic length scales may be parametrically longer than the lattice scales. 
We may also generically expect the compressibility coefficient $K$ in the Nonlinear $\sigma$-model in the main text to be of order $\nu(E_F) |\Delta(0)|^2$ where $\Delta(0)$ is the zero-temperature chiral-condensate gap, and the phase stiffnesses $J/K,J'/K \sim v_F^2/q_\star^2$ though this should be investigated more thoroughly.

\end{document}